\documentclass[prl,twocolumn,showpacs,preprintnumbers,amsmath,amssymb]{revtex4}
\usepackage{graphicx}% Include figure files
\usepackage{dcolumn}% Align table columns on decimal point
\usepackage{bm}% bold math
\begin{document}
\title{Anisotropic magnetoresistance and anisotropic tunneling magnetoresistance \\due to quantum interference in ferromagnetic metal break junctions}
\author{Kirill I. Bolotin}
\author{Ferdinand Kuemmeth}
\author{D. C. Ralph}
%\email{ralph@ccmr.cornell.edu}
\affiliation{Laboratory of Atomic and Solid State Physics, Cornell University, Ithaca, NY 14853 USA}
\date{\today}

\begin{abstract}
We measure the low-temperature resistance of permalloy break junctions as a function of contact size and the magnetic field angle, in applied fields large enough to saturate the magnetization.  For both nanometer-scale metallic contacts and tunneling devices we observe large changes in resistance with angle, as large as 25\% in the tunneling regime. The pattern of magnetoresistance is sensitive to changes in bias on a scale of a few mV.  We interpret the effect as a consequence of conductance fluctuations due to quantum interference.
\end{abstract}
%\keywords{Suggested keywords}%Use showkeys class option if keyword
       
\maketitle
The magnetoresistance properties of nanometer-scale magnetic devices can be quite different from those of larger samples.  One aspect of this difference has been explored extensively in previous experiments -- the resistance of magnetic domain walls created when the magnetic moment direction in one magnetic electrode is rotated relative to the moment in a second electrode \cite{levy,garcia,hua,viret,gabureac,yang,egelhoff,keane,bolotin}. Here we focus on a different aspect of the physics of magnetoresistance in nanoscale magnetic contacts -- the anisotropic magnetoresistance (AMR) that arises when the magnetization throughout a device is rotated uniformly so as to change the angle between the direction of current flow and the magnetic moment. 
Our measurements are motivated by predictions of increased AMR for atomic-sized ballistic conductors \cite{velev} and indications of enhanced AMR in Ni contacts \cite{keane}.  By making detailed studies of resistance as a function of field angle using mechanically-stable permalloy contacts, we show that the size of the AMR signal at low temperature can increase dramatically as the contact cross section is narrowed to the nanometer-scale regime.  Even more strikingly, we find that point contacts which are completely broken, so as to enter the tunneling regime, also exhibit a tunneling anisotropic magnetoresistance effect (TAMR) as large as 25\% when the magnetic-moment directions in the two contacts are rotated together while remaining parallel. 

\begin{figure}
\includegraphics{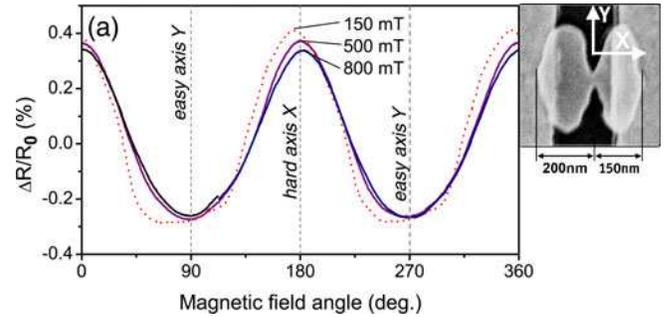}
\caption{(a) Zero-bias differential resistance vs.\ angle of applied magnetic field at different field magnitudes at 4.2 K, illustrating bulk AMR for a constriction size of $30 \times  100$ nm$^2$ and resistance $R_0 =$ 70 $\Omega$ (device A). (b) SEM micrograph of a typical device.}
\label{fig1}
\end{figure}

Magnetostriction and magnetostatic forces can alter the geometry of nanoscale junctions as the magnetic field is varied, and produce artifacts in the resistance, so experiments must be designed to minimize these effects \cite{gabureac,yang,egelhoff}. For this reason, our contacts are firmly attached to a non-magnetic silicon substrate and are measured entirely at low temperature to suppress thermally-driven surface diffusion of metal atoms.  Similar structures have proven \cite{keane,bolotin} to be much more mechanically-stable than previous samples which were measured at room temperature. 
We fabricate our devices using aligned steps of electron beam lithography to first pattern 20-nm-thick gold contact pads and then 30-nm-thick magnetic permalloy (Py = Ni$_{\textrm{80}}$Fe$_{\textrm{20}}$) point contacts \cite{bolotin}. Each contact consists of two elongated electrodes which are connected by a 100-nm-wide bridge (Fig.~\ref{fig1}(b)). The magnetic field $\bf{B}$ is applied using a 3-coil vector magnet capable of 0.9 T in any direction and up to 7 T along one axis (the x axis, defined below) with the other two coils turned off. The differential resistance $R=dV/dI$ at voltage bias $V$ is measured using a lock-in amplifier with an excitation voltage small enough not to broaden the data; a total of 46 devices were studied.  

Measurements are performed as follows: we first cool the samples to 4.2 K and narrow the size of the bridge connecting the two magnetic electrodes by using actively controlled electromigration \cite{strachan}. When the desired cross-section is reached (as determined by the sample's $R$) we stop the electromigration process and measure $R$ at 4.2 K while rotating the magnetic field in the plane of the sample at fixed magnitude. Then the same procedure is repeated to achieve smaller device cross-sections and larger $R$. As a result we can examine magnetic properties of each device as a function of the bridge size, down to the atomic scale and into the tunneling regime \cite{bolotin}.

The resistances of all devices before electromigration ($\approx 70~\Omega$ at 4.2 K) exhibit a small periodic dependence on the field direction ($\sim 1\%$, Fig.~\ref {fig1}(a)). This is a signature of the bulk anisotropic magnetoresistance (AMR), which for a polycrystalline sample may be written as
$\Delta R \propto \cos^{2}(\theta)$,
where $\theta$ is the angle between the current and the magnetization $\bf{M}$ \cite{mcguire}. The resistance of our devices before electromigration is maximal for $\bf{B}$ applied in the x-direction (Fig.~\ref{fig1}(b)), parallel to the current.  We measure $\theta$ relative to this direction.

\begin{figure}
\includegraphics{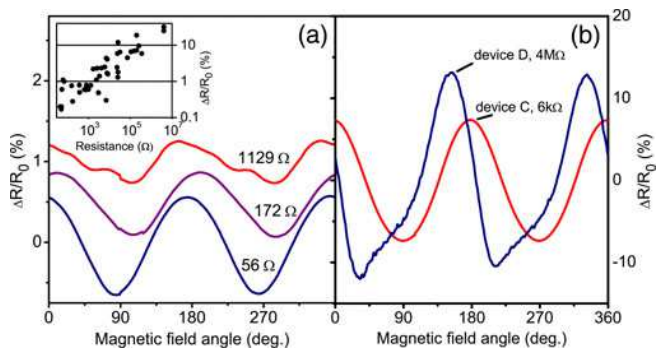}
\caption{(a) Evolution of AMR in device B as its resistance $R_0$ is increased from 56~$\Omega$ to 1129~$\Omega$. (b) AMR for a $R_0 =$ 6 k$\Omega$ device (device C) exhibiting 15\% AMR, and a $R_0 =$ 4 M$\Omega$ tunneling device (device D), exhibiting 25\% TAMR.  All measurements were made at a field magnitude of 800 mT at 4.2 K. Inset: AMR magnitude as a function of $R_0$ for 12 devices studied into the tunneling regime.}
\label{fig2}
\end{figure}

Because the AMR depends on the orientation of the magnetization, it is important to ensure that the sample is magnetized uniformly and always remains saturated in the direction of the applied field. We estimated the distribution of magnetization within our sample using the OOMMF code \cite{oommf}.   Such modeling suggests that applying 800 mT effectively saturates the nanoscale magnetic electrodes for all directions in plane: the average $\bf{M}$ follows $\bf{B}$ to within 2$^\circ$ and the RMS fluctuation in the angle of magnetization across the sample is  $\sigma_{\bf{M}} < 4^\circ$. To check this experimentally, we fit our 800 mT data to $\Delta R \propto \cos^{2}(\theta)$, and we found that the RMS deviation of the magnetization angle indicated by the fit was  $\sigma_{\bf{M}} < 5^\circ$.  We observe that the applied field becomes insufficient to fully saturate $\bf{M}$ below approximately 200 mT, at which point $\bf{M}$ departs from the field direction toward the easy axis of the sample (Fig.~\ref{fig1}(a), dotted curve). We performed similar studies also for samples in the tunneling regime and for near-atomic-sized contacts. In addition, we performed sweeps to 7 Tesla along the hard in-plane axis (x axis) for one sample having $R=$ 3~k$\Omega$  in the metallic range and two samples in the tunneling regime 200, 400~k$\Omega$.  (Device E with $R=$ 2.6~k$\Omega$ was measured to 3.5 T.)  In all cases we found that 0.8 T in-plane magnetic fields were sufficient to saturate the resistance.

As the cross section of the device is narrowed for samples with $R <$ 500 $\Omega$, both the phase and the amplitude of the AMR can change, but the AMR remains small and retains its $\cos^2(\theta)$ dependence (Fig.~\ref{fig2}(a), $R=$ 172 $\Omega$).  The changes in phase and amplitude may be a result of changes in sample geometry during electromigration.  Scanning electron microscopy studies show changes large enough to alter the direction of current flow in the junction.

\begin{figure}
\includegraphics{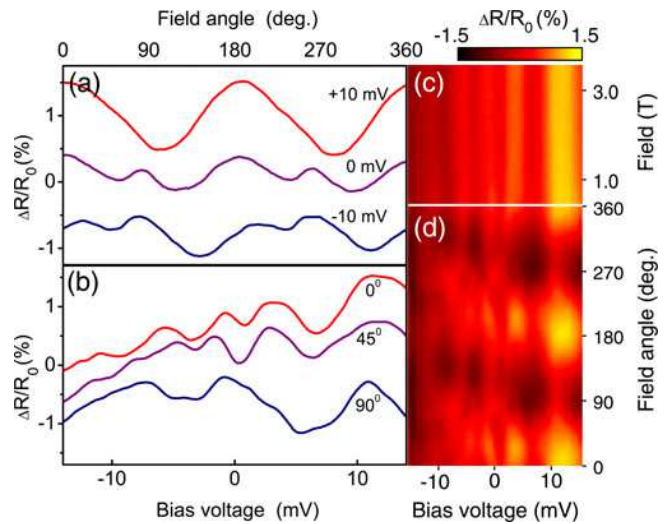}
\caption{Variations of $R = dV/dI$ at 4.2 K in a sample with average zero-bias $R_0 =$ 2.6 k$\Omega$ (device E). (a) $R$ vs.\ field angle at different bias voltages ($|B|=$ 800 mT). (b) Dependence of $R$ on $V$ at different fixed angles of magnetic field ($|B|=$ 800 mT). The curves in (a) and (b) are offset vertically. (c) $R$ as a function of $V$ and magnetic field strength, with field directed along the x axis. $R$ does not have significant dependence on the magnitude of $B$. (d) $R$ as a function of $V$ and $\theta$, for $|B|=$ 800 mT.}
\label{fig3}
\end{figure}

As the cross-section is reduced further, to the regime where $R$ is larger than several hundred $\Omega$, the angular dependence of the AMR for some samples (Fig.~\ref{fig2}(a)) can become more complicated than the simple $\cos^2(\theta)$ form. In addition, we find that devices with $R$ larger than $\sim$1 k$\Omega$ generally exhibit larger AMR. Figure~\ref{fig2}(b) shows a 6 k$\Omega$ device with an AMR of 14\% (device C), more than 50 times the value for this device before electromigration. Even for samples in the tunneling regime ($R> h/e^2 \approx $ 26 k$\Omega$) we continue to measure large values of AMR, as large as 25\% in a 2 M$\Omega$ sample (Fig.~\ref{fig2}(b), device D). The dependence of the AMR on sample resistance is shown in Fig.~\ref{fig2}, Inset.

\begin{figure}
\includegraphics{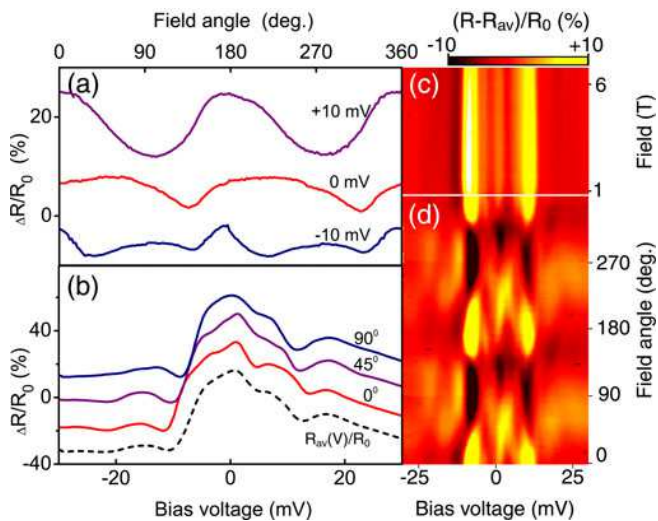}
\caption{Variations of $R = dV/dI$ at 4.2 K in a sample with average zero-bias $R_0 =$ 257~k$\Omega$ (device F), in the tunneling regime.(a) $R$ vs.\ field angle at different bias voltages ($|B|=$ 800 mT). (b) Dependence of $R$ on $V$ at different fixed angles of magnetic field ($|B|=$ 800 mT). The curves in (a) and (b) are offset vertically. (c) $R(V)-R_{\rm av}(V)$ as a function of $V$ and magnetic field strength, with field directed along the x axis. $R_{\rm av}(V)$, the average of $R(V)$ over angle (shown in (b)), is subtracted to isolate angular-dependent variations. (d) $R(V)-R_{\rm av}(V)$ as a function of bias voltage and magnetic field angle, for $|B|=$ 800 mT.  
}
\label{fig4}
\end{figure}

We can gain insight into the mechanism behind the large AMR and TAMR effects from their dependence on bias voltage.  There are significant changes in the angular dependences of $dV/dI$ for voltages differing by just a few mV (Figs.~\ref{fig3}(a),\ref{fig4}(a)).  
Moreover, at fixed field angle, $dV/dI$ also exhibits reproducible fluctuations as a function of $V$ (Figs. \ref{fig3}(b),\ref{fig4}(b)). These fluctuations depend only on the angle of the applied field, not on its strength (Figs. \ref{fig3}(c),\ref{fig4}(c)). For both the metallic and tunneling samples the size of the AMR effect is similar to the magnitude of fluctuations in $dV/dI$ as a function of $V$.

Before discussing other mechanisms, we consider the possibility of artifacts due to magnetostriction and magnetostatic forces.  Neither of these effects should produce smooth fluctuations in $R$ as a function of small changes in $V$.  Furthermore, for samples with $R$ near that of a single quantum channel, these effects are known to cause atomic rearrangements manifested as irreproducible jumps in $R$ \cite{gabureac}, a feature not seen in any of our data. 
We can estimate the consequences of magnetostriction in the tunneling regime by assuming that the magnetostriction constant in Py is $\lambda_s<10$ ppm and the length of any suspended region in our device is $<10$ nm, so that any displacement is $<$ 10~fm.  Applying the Simmons formula for tunneling \cite{wolf} with a work function $<$ 5~eV, the change in $R$ due to this displacement would be $<$
0.4\%, more than 50 times smaller than the AMR we observe for tunneling devices. Magnetostatic forces would give changes in $R$ of the opposite sign than we measure for many samples. We conclude that neither magnetostriction nor magnetostatic effects can account for our enhanced AMR signals.

Fluctuations in $R$ as a function of $V$, similar to those we measure, have been observed previously in non-magnetic samples and are understood to be a signature of mesoscopic quantum interference of scattered electron waves \cite{leestone}.  For diffusive metal samples with a characteristic size similar to the dephasing length, the magnitude of the fluctuations has a universal scale when expressed in terms of the conductance ($G = dI/dV = 1/R$): $\Delta G \sim e^2/h$ in nonmagnetic samples with weak spin-orbit scattering and $\Delta G \sim 0.4 e^2/h$ in magnetic samples with spin-orbit scattering \cite{leestone}.  However, the conductance fluctuations in non-magnetic point-contact devices with a contact radius less than the elastic mean free path $l_e$ have smaller, non-universal magnitudes \cite{holweg,ralph,ludoph}. The average magnitude of the fluctuations that we measure in samples with $R =$ 1-14 k$\Omega$ is 0.1 $e^2/h$. Conductance fluctuations as a function of $V$ have also been observed previously for small non-magnetic tunnel junctions \cite{vanoud}, and are understood to be a consequence of mesoscopic fluctuations in the local density of electronic states of a disordered sample. Because the variations that we measure in $R$ as a function of $\theta$ have a magnitude similar to the fluctuations as a function of $V$, we propose that the dominant process giving rise to enhanced AMR and TAMR in our samples is mesoscopic interference, as well. 

Unlike previous measurements in non-magnetic devices \cite{holweg,vanoud}, we do not observe fluctuations as a function of the magnitude of magnetic field up to at least 7 T (Figs. \ref{fig3}(c),\ref{fig4}(c)), only as a function of $\theta$.  Based on the data, we estimate that the correlation scale for fluctuations as a function of field magnitude must be $B_c > 20$ T.  We therefore conclude that our AMR and TAMR cannot be due directly to the magnetic field affecting the Aharonov-Bohm phase of the electrons; the maximum change in total field through the sample upon rotating the magnetization by $90^{\circ}$ at 0.8 T is only $\sqrt{2}(\mu_0 M_s + 0.8$ T$) \sim 2.7$ T, where $\mu_0 M_s = 1.1$ T is the magnetization for permalloy. However, an alternative mechanism was recently proposed by Adam {\em et~al.}\ \cite{piet}, that rotation of the magnetization direction in ferromagnets may alter quantum interference because it is coupled to the electrons' orbital motion via spin-orbit scattering. As a result, mesoscopic fluctuations in the conductance of magnetic metal samples and in the local density of states of magnetic tunneling devices can be expected to occur as a function of the magnetization orientation. 

The theory of Adam {\em et~al.}\ \cite{piet} was solved for the case of diffusive samples, and therefore one should not expect it to be quantitative for our point contacts.  Nevertheless, we will compare the results of this theory to our measured correlation scales, to test whether the mechanism of Adam {\em et~al.}\ might provide a reasonable qualitative explanation.  
The voltage correlation scale for our data is $V_c \approx$ 1-2~mV, approximately equal to the limit set by thermal broadening at 4.2 K.
The zero-temperature energy correlation scale $E_c$ can be calculated by the formalism in ref. \cite{piet} to be 
\begin{equation}
E_c = (E^{\uparrow}_{T}  \tau^{\uparrow}_\bot + E^{\downarrow}_T  \tau^{\downarrow}_{\bot})/(\tau^{\downarrow} _{\bot} + \tau^{\uparrow}_{\bot}),
\label{eq2}
\end{equation}
where $E_{T}^{\uparrow}$ and $E_{T}^{\downarrow}$ are the Thouless energies for spin-up and spin-down s,p-band electrons and $\tau^{\uparrow}_{\bot}$ and $\tau^{\downarrow}_{\bot}$ are spin-flip spin-orbit scattering times \cite{saar}.  
In permalloy, because of the contribution of the minority-electron d-band, the density of states at the Fermi level for minority electrons $\nu^{\downarrow}$ is much greater than for majority electrons, so by Fermi's golden rule we can estimate $\tau_{\bot}^{\downarrow} \propto (\nu^{\uparrow})^{-1} \gg \tau^{\uparrow}_{\bot} \propto (\nu^\downarrow)^{-1}$ and $E_{T}^{\uparrow} \tau^{\uparrow}_{\bot}\approx E_{T}^{\downarrow} \tau^{\downarrow}_{\bot} \propto (\nu^\uparrow \nu^\downarrow)^{-1}$. Eq.~(\ref{eq2}) then takes a simple form, $E_c \sim 2 E_{T}^{\downarrow} \sim 2 \pi^2 \hbar v_F l_{e}^{\downarrow} /3 L^2_{\phi}$, where $v_F = 0.2 \times 10^6$~m/s is the Fermi velocity in Py \cite{petrovykh}, $l_{e}^{\downarrow} \sim 0.6$~nm is the elastic mean free path for minority electrons \cite{petrovykh,gurney}, and $L_{\phi}$ is the dephasing length. Assuming that the voltage correlation scale $V_c \sim \max \{k_B T/e, E_c/e \}$, we find a rough lower limit on the dephasing length, $L_{\phi} \gtrsim 16$~nm. 
If $L_{\phi}$ is close to this value, then the magnetic field correlation scale $B_c \sim \Phi_0/ L^2_{\phi} \sim 16$~T, in reasonable agreement with our estimate from the field dependence. The formalism in ref. \cite{piet} can also be used to predict the correlation angle for the fluctuations \cite{saar}: 
\begin{equation}
\theta_c \sim \sqrt{\frac{2}{\hbar} \left(E^{\uparrow}_{T} \tau^{\uparrow}_{\bot} + E^{\downarrow}_{T} \tau^{\downarrow} _{\bot}\right)/\left(2 + \frac{\tau^{\uparrow}_{\bot}}{\tau^{\uparrow}_{\|}} +\frac{\tau^{\downarrow}_{\bot}}{\tau^{\downarrow}_{\|}}\right)},
\label{eq3}
\end{equation}
where $\tau^{\uparrow}_{\|}$ and $\tau^{\downarrow}_{\|}$ are mean free times for spin-conserving spin-orbit scattering. Employing golden-rule assumptions similar to those we used above: $\tau^\uparrow_\bot$, $\tau^\downarrow_\| \propto (\nu^\downarrow)^{-1}$ and $\tau^\downarrow_\bot$,  $\tau^\uparrow_\| \propto (\nu^\uparrow)^{-1}$, we find $\theta_c \sim 2 (\tau^\downarrow_\| E^\downarrow_T /\hbar)^{1/2} \sim \frac{2 \pi}{\sqrt{3}} (\tau^\downarrow_\| / \tau^{\downarrow}_{e})^{1/2} l_e^{\downarrow}/L_{\phi}$.  With the approximations $\tau^\downarrow_\| \sim 2 \tau^\uparrow_\bot$ \cite{saar}, $\tau^{\uparrow}_{\bot} \sim (5.5 {\rm nm})/v_F$ \cite{steenwyk}, our estimate for $\theta_c$ is $\sim 0.6$~radians.
Considering the rough nature of the approximations, we consider this to be in good agreement with our measurements -- typically we see one or two oscillations in $dV/dI$ as a function of $\theta$ over the relevant range of 0 to $\pi$ radians.  (By inversion symmetry, $R$ at $V$=0 must be unchanged upon rotation by $\pi$.)  

Large TAMR signals, qualitatively similar to our results in the tunneling regime, were also reported recently in (Ga,Mn)As magnetic semiconductor tunnel junctions \cite{gould}.  However, the mechanism proposed to explain the (Ga,Mn)As measurements is a band-structure effect by which the {\em bulk} density of states depends on magnetization angle \cite{gould,shick}.  This is fundamentally distinct from our proposal that TAMR effects in nanoscale metal devices are due to mesoscopic fluctuations in the {\em local} density of states.  As already noted in ref. \cite{piet}, mesoscopic fluctuations as a function of magnetization angle may be relevant in describing another recent experiment \cite{drago}, which was originally analyzed in terms of the motion of magnetic domain walls.  

In summary, we have measured the AMR of ferromagnetic metal contacts at low temperature as a function of their size, over the range from large ($100\times30$ nm$^2$) cross sections to atomic-scale point contacts and into the tunneling regime. For metallic devices with $R$ larger than $\sim 1$\ k$\Omega$ we observe AMR effects larger than in bulk devices, with an angular variation that can deviate from the sinusoidal bulk dependence, and which are associated with fluctuations in $R$ of similar magnitude as a function of $V$. Similar effects are also seen in magnetic point-contact tunneling devices. We propose that these large AMR and TAMR effects are the result of mesoscopic quantum interference which depends on the orientation of the magnetization, leading to fluctuations of conductance and the spin-dependent local density of states. These fluctuations should affect a broad variety of nanoscale devices that contain magnetic components, producing strong perturbations in measurements of low-temperature spin-dependent transport. 

Note added: M. Viret et al.\ have recently posted related, but contrasting results \cite{viret_new}.

We thank Saar Rahav and Piet Brouwer for discussions and Eric Smith for experimental help. This work was funded by the NSF (DMR-0244713 and through use of the Cornell NanoScale Facility/NNIN) and by the ARO (DAAD19-01-1-0541).

\end{document}